\documentclass[12pt]{article}
\begin{document}
\bibliographystyle{unsrt}
\title{\bf Vlasov Equation Of Plasma In Magnetic Field}
\author{Biao Wu
\\ Department of Physics
\\ The University of Texas at Austin, Austin, TX-78712, USA}
\maketitle
\begin{abstract}
The linearized Vlasov equation for a plasma system 
in an external constant magnetic field and the corresponding linear 
Vlasov operator are studied. The solution of the Vlasov equation is found 
by the resolvent method. The spectrum and eigenfunctions of the Vlasov 
operator are also found. The spectrum of this operator consists of two parts: 
one is continuous and real; the other is discrete and complex. Interestingly, 
the real eigenvalues are uncountably infinitely degenerate, 
which causes difficulty solving this initial value problem by using the conventional
eigenfunction expansion method. It also breaks the natural relation between 
the eigenfunctions and the resolvent solution that the eigenfunctions can be considered 
as the coefficients of $e^{-i\omega t}$ in the Laplace (or resolvent) solution.
\end{abstract}
\newpage
\section{Introduction}
In plasma physics, many interesting phenomena such as plasma oscillations, 
instabilities and Landau dampings can be studied and understood through a 
very simple model of plasma. In this model, ions are assumed to be motionless 
and form a positively charged background and the collisions between electrons 
are neglected. In 1945, Vlasov proposed a nonlinear equation, now known as Vlasov 
equation, to describe this model~\cite{Vla}. Since then, a lot of effort 
has been made to study this equation. When there is no external field, one usual way 
to treat this problem is to first linearize the equation by assuming the system is 
very close to the equilibrium state, then reduce it to a one-dimensional equation.
Many interesting phenomena can be understood through this simple one-dimensional
linearized Vlasov equation.\cite{Lan, Bal, Ich}\\

In this paper, I shall study this system in an external constant magnetic field.
After the Vlasov equation is linearized, two methods, the eigenfunction expansion
method and the resolvent method, are tried to solve the equation. The resolvent method 
is proved to be successful. The eigenfunction expansion method is, surprisingly, 
not successful even after all the eigenfunctions of the corresponding linear operator, 
Vlasov operator, are worked out. The difficulty is caused by the fact that the real 
eigenvalues are uncountably infinitely degenerate. \\

It has been studied by Case~\cite{Cas} and Arthur et al~\cite{Art}
the explicit relation between the Laplace transform (or the resolvent) approach and the 
eigenfunction expansion approach to the one-dimensional Vlasov equation with 
no external field. 
Especially, the latter demonstrated how to construct the van Kampen-Case
modes~\cite{Cas, van}
through the resolvent solution. Their results show that the 
van Kampen-Case modes are, loosely speaking,
 just the coefficients of $e^{i\omega t}$ in the resolvent 
(or Laplace) solution. This is no longer
true when we don't reduce the Vlasov equation to a one-dimensional problem or 
simply we can't because of the existence of external fields. The eigenmodes 
(or eigenfunctions) corresponding to the real eigenvalues presented in Section 3 
can not be constructed with the resolvent solution in Section 5, which is caused 
by the infinite degeneracy mentioned above.\\
 
The plan of this paper is as follows. In Section 2, the derivation of the 
linearized Vlasov equations and its related linear Vlasov operators are sketched, 
primarily to introduce notations. The eigenfunctions are presented and discussed 
in Section 3 without detailed calculations. Next, the related adjoint problem is 
discussed briefly and the orthogonal relation between the original eigenfunctions 
and the adjoint eigenfunctions is proved. In Section 5, the exact solution is given 
by using the resolvent method.

\section{Vlasov Equations and Vlasov Operators}
This simple model of plasma is described by the Vlasov equation 
\begin{equation}
{\partial\over \partial t}\rho(\vec{x},\vec{v},t)= 
-\vec{v}\cdot{\partial\over \partial \vec{x}}\rho(\vec{x},\vec{v},t)
-{\vec{F}\over m_e}\cdot {\partial\over \partial\vec{v}}\rho(\vec{x},\vec{v},t)
\end{equation}
where $\rho(\vec{x},\vec{v},t)$ is the single distribution function of electrons and 
$\vec{F}$ is the force felt by a single electron. In this paper, the
system is in an 
external constant magnetic field $\vec{B}_{0}$, so
\begin{equation}
\vec{F}=-e\vec{v}\times \vec{B}_{0}-e\vec{E}
\end{equation}
where $\vec{E}(\vec{x},t)=\displaystyle e{\partial\over \partial \vec{x}}\int 
{\rm d}\vec{x}'{\rm d}\vec{v}'{\rho(\vec{x}',\vec{v}',t)\over 
|\vec{x}-\vec{x}'|} $
is the electric field at position $\vec{x}$ and time $t$ generated 
by the electron system itself.
If the system is just slightly away from an equilibrium state $f_0(\vec{v})$
\begin{equation}
\rho(\vec{x},\vec{v},t)=n_0 f_0(\vec{v})+f(\vec{x},\vec{v},t)
\end{equation}
where $n_0$ is the average density and $f\ll n_0 f_0$, then it can be 
described very well by the linearized Vlasov equation which is  
\begin{equation}
{\partial f\over \partial t}=
-\vec{v}\cdot{\partial f\over \partial \vec{x}}+{e\over m_e}(\vec{v}\times \vec{B}_{0})
\cdot{\partial f\over \partial \vec{v}}+ 
{e^2 n_0\over m_e}[{\partial\over \partial \vec{x}}\int
{\rm d}\vec{x}'{\rm d}\vec{v}'{f\over
|\vec{x}-\vec{x}'|}]\cdot {\partial f_0\over \partial
\vec{v}}
\end{equation}
Taking the Fourier transform
\begin{equation}
f(\vec{x},\vec{v},t)=\sum_{\vec{k}\neq 0}f_{\vec{k}}(\vec{v},t)
\exp{({\rm i}\vec{k}\cdot\vec{x})}
\end{equation}
we have for the $\vec{k}$-component function $f_{\vec{k}}(\vec{v},t)$
\begin{equation}
\begin{array}{rl}
\displaystyle {\rm i}{\partial\over \partial t}f_{\vec{k}}(\vec{v},t)
=&\displaystyle
\vec{k}\cdot\vec{v}f_{\vec{k}}(\vec{v},t)+{\rm i}{e\over m_e}(\vec{v}\times\vec{B}_{0})
\cdot{\partial\over \partial \vec{v}}f_{\vec{k}}(\vec{v},t)\\~\\

&\displaystyle -{\omega^2_p\over k^2}\vec{k}\cdot{\partial\over 
\partial \vec{v}}f_{0}(\vec{v})\int{\rm d}\vec{v}'f_{\vec{k}}(\vec{v}',t)
\end{array}
\end{equation}
where $\omega_p=(4\pi e^2 n_0/ m_e)^{1/2}$ is the plasma frequency. 
To simplify the notation, let us drop the index $\vec{k}$ and let
\begin{equation}
\eta(\vec{v})={\omega^2_p\over k^2}\vec{k}\cdot{\partial\over 
\partial \vec{v}}f_{0}(\vec{v})
\end{equation}
then equation (6) becomes
\begin{equation}
{\rm i}{\partial\over \partial t}f(\vec{v},t)=
\vec{k}\cdot\vec{v}f(\vec{v},t)+{\rm i}{e\over m_e}(\vec{v}\times\vec{B}_{0})
\cdot{\partial\over \partial \vec{v}}f(\vec{v},t)-\eta(\vec{v})
\int{\rm d}\vec{v}'f(\vec{v}',t)
\end{equation}
This is the equation I shall study mostly in this paper. Consequently, 
the corresponding linear Vlasov operator ${\mathcal K}$ is defined as
\begin{equation}
{\mathcal K}f(\vec{v})=\vec{k}\cdot\vec{v}f(\vec{v})+{\rm i}{e\over m_e}
(\vec{v}\times\vec{B}_{0})\cdot{\partial\over \partial \vec{v}}f(\vec{v})
-\eta(\vec{v})\int{\rm d}\vec{v}'f(\vec{v}')
\end{equation}
This is an integro-differential, unbounded and non-self-adjoint 
linear operator. In terms of ${\mathcal K}$, equation (8) can be put into a 
compact form
\begin{equation}
{\rm i}{\partial\over \partial t}f(\vec{v},t)={\mathcal K}f(\vec{v},t) 
\end{equation}
The equation for the case where $\vec{B}_{0}=0$ is  
\begin{equation}
{\rm i}{\partial\over \partial t}f(\vec{v},t)={\mathcal K_0}f(\vec{v},t) 
\end{equation}
with 
\begin{equation}
{\mathcal K_0}f(\vec{v})=\vec{k}\cdot\vec{v}f(\vec{v})
-\eta(\vec{v})\int{\rm d}\vec{v}'f(\vec{v}')
\end{equation}
Integrating over the two components of $\vec{v}$ perpendicular to $\vec{k}$, 
we can reduce (11) to a one-dimensional problem 
\begin{equation}
{\rm i}{\partial\over \partial t}{\bar f}(v,t)={\bar {\mathcal K}}_0{\bar f}(v,t)
\end{equation}
with
\begin{equation}
{\bar {\mathcal K}}_0 {\bar f}(v)=kv{\bar f}(v)-{\bar \eta}(v)\int_{-\infty}^{\infty}
{\rm d}v'{\bar f}(v')
\end{equation}
Here $v$ is the component of $\vec{v}$ along the direction of 
$\vec{k}$ and $k$ is the magnitude of $\vec{k}$. The bar over
functions indicates that they have been integrated over the two  
components of $\vec{v}$ perpendicular to $\vec{k}$. This convention is assumed 
throughout this paper. Equation (13) is the one which has been studied thoroughly 
by van Kampen, Case and many others~\cite{Lan,Cas,Art,van,Bac}. \\

Formally, equations (10), (11) and (13) are just the same as the Schr{\" o}dinger
equation. But the Hamiltonian operator appearing in the Schr{\" o}dinger equation
is self-adjoint and bounded from the below while ${\bar {\mathcal K}}_0$,
${\mathcal K_0}$ and ${\mathcal K}$ are unbounded and not self-adjoint. 
These are all typical
equations arising in linear evolution systems. There are many standard methods to 
study this class of linear equations. For example, equation (13) can be solved
by the Laplace transform~\cite{Lan, Cas}, the resolvent method~\cite{Art} and 
the eigenfunction expansion method~\cite{Cas, van}. I shall try to use the 
eigenfunction expansion method and the resolvent method to solve equation (10) and (11). 
  
\section{Eigenfunctions and Eigenvalues}
In this section, I shall present the eigenfunctions and spectra of operators 
${\bar {\mathcal K}}_0$, 
${\mathcal K_0}$ and ${\mathcal K}$. These three operators are similar in many respects 
as they are supposed to be. Their spectra are the same: continuous real eigenvalues, discrete
real eigenvalues and discrete complex eigenvalues. Since the existence of discrete real
eigenvalues depends on the choice of the equilibrium function $f_0(\vec{v})$, 
I shall only consider, for simplicity, the case where the discrete real eigenvalues don't 
exist. Also, their eigenfunctions are similar: most of them are singular containing 
$\delta$ functions. However, there is one big difference: the real eigenvalues of 
${\mathcal K_0}$ and ${\mathcal K}$ are infinitely degenerate while the eigenvalues 
of ${\bar {\mathcal K}}_0$ are not degenerate. This infinite degeneracy causes the 
difficulty expanding functions in terms of these eigenfunctions and relating these 
eigenfunctions with the resolvent solutions.\\

\noindent{\large \it Operator ${\bar {\mathcal K}}_0$}:\\

\noindent The eigenfunctions of ${\bar {\mathcal K}}_0$ were first constructed 
by van Kampen~\cite{van} then completed by Case~\cite{Cas} and are now widely known as 
van Kampen-Case modes. Their results can be summarized as follows.\\

\noindent The eigenequation for eigenvalue $z$ is 
\begin{equation}
{\bar {\mathcal K}}_0 {\bar g}_{z}(v)=kv{\bar g}_{z}(v)-
{\bar \eta}(v)\int_{-\infty}^{\infty}{\rm d}v'{\bar g}_{z}(v')=z{\bar g}_{z}(v) 
\end{equation}
It is linear so the functions can be normalized as
\begin{equation}
\int_{-\infty}^{\infty}{\rm d}v{\bar g}_{z}(v)=1
\end{equation}
When the eigenvalue is real, say, it is $\nu$, the eigenfunction is
\begin{equation}
{\bar g}_{\nu}(v)={\mathcal P}{{\bar \eta}(v)\over kv-\nu}+{\bar \lambda}(v)
\delta(kv-\nu)
\end{equation}
Here ${\mathcal P}$ means the principal value integration.
The normalization condition (16) requires
\begin{equation}
{1\over k}{\bar \lambda}({\nu\over k})=1-
\int{\rm d}v{\mathcal P}{{\bar \eta(v)}\over kv-\nu}
\end{equation}
The discrete complex eigenvalues $\nu_j$ ($j=1,2,...,m_0$) are determined by 
\begin{equation}
\epsilon_0(z)=1-\int_{-\infty}^{\infty}{\rm d}v{{\bar \eta}(v)\over kv-z}
=1-\int{\rm d}\vec{v}{\eta(\vec{v})\over \vec{k}\cdot\vec{v}-z}=0
\end{equation}
The corresponding eigenfunction is
\begin{equation}
g_j(v)=g_{\nu_j}(v)={{\bar \eta}(v)\over kv-\nu_j}
\end{equation}
which satisfies normalization condition (16). 
Eigenfunctions (17) and (20) are the famous van Kampen-Case modes. It is clear 
that all the eigenvalues of ${\bar {\mathcal K_0}}$ are 
not degenerate when the van Kampen-Case modes are  
normalized according to (16). \\

\noindent{\large \it Operator ${\mathcal K_0}$}:\\

\noindent The eigenequation is 
\begin{equation}
{\mathcal K_0}g_{z}(\vec{v})=\vec{k}\cdot\vec{v}g_{z}(\vec{v})
-\eta(\vec{v})\int{\rm d}\vec{v}'g_{z}(\vec{v}')=z g_{z}(\vec{v})
\end{equation}
As this operator is linear, we set the normalization condition
\begin{equation}
\int{\rm d}\vec{v}g_{z}(\vec{v})=1
\end{equation}
The eigenfunction corresponding to a real eigenvalue $\nu$ is 
\begin{equation}
g_{\nu}(\vec{v})={\mathcal P}{\eta(\vec{v})\over \vec{k}\cdot\vec{v}-\nu}
+\lambda(\vec{v})\delta(\vec{k}\cdot\vec{v}-\nu)
\end{equation}
The normalization condition (22) requires that $\lambda(\vec{v})$ satisfy
\begin{equation}
\int{\rm d}\vec{v}\lambda(\vec{v})\delta(\vec{k}\cdot\vec{v}-\nu)=1-
\int{\rm d}\vec{v}{\mathcal P}{\eta(\vec{v})\over \vec{k}\cdot\vec{v}-\nu}
\end{equation}
It is very clear here that there are infinitely many choices for $\lambda(\vec{v})$
to satisfy the above condition. For example, all functions 
$\lambda(\vec{v})=\alpha(v_1, v_2){\bar \lambda}(v)$ satisfy (24) as long as
\begin{equation}
\int_{-\infty}^{\infty}{\rm d}v_1\int_{-\infty}^{\infty}{\rm d}v_2
\alpha(v_1,v_2) = 1
\end{equation}
where $v_1$ and $v_2$ are the two components of $\vec{v}$ perpendicular to $\vec{k}$.
Obviously, the choices of $\alpha(v_1, v_2)$ are uncountably 
infinite, which means there are infinitely many corresponding 
eigenfunctions for any real eigenvalue $\nu$.
In other words, the degeneracy of every real eigenvalue is 
uncountably infinite.\\

The complex eigenvalues of ${\mathcal K_0}$ are the same as 
${\bar{\mathcal K}}_0$, $\nu_j$ ($j=1,2,...,m_0$), the zeroes of $\epsilon_0 (z)$,
whose corresponding eigenfunctions are
\begin{equation}
g_j(\vec{v})=g_{\nu_j}(\vec{v})={\eta(\vec{v})\over \vec{k}\cdot\vec{v}-\nu_j}
\end{equation}
~\\

\noindent {\large \it Operator ${\mathcal K}$}:\\

\noindent I shall present the results 
directly without going into the detailed calculations.
The method of computing eigenfunctions is very similar to the one given in the Appendix.
First, let us set up a coordinate system and introduce some notations:
\begin{equation}
\vec{B}_0=B_0\hat{z}~~,~~~~~~~~\vec{k}=k_{\bot}\hat{x}+k_{\|}\hat{z}
\end{equation}
\begin{equation}
\vec{v}=v_{\bot}\cos\theta\hat{x}+v_{\bot}\sin\theta\hat{y}+v_{\|}\hat{z}
\end{equation}
Thus the operator ${\mathcal K}$ becomes
\begin{equation}
{\mathcal K}f(\vec{v})=(k_{\bot}v_{\bot}\cos\theta+k_{\|}v_{\|})f(\vec{v})-{\rm i}\omega_0
{\partial \over \partial\theta}f(\vec{v})-\eta(\vec{v})\int{\rm d}\vec{v}'f(\vec{v}')
\end{equation}
where $\omega_0=eB_0/m_e$ is the cyclotron frequency. The same normalization condition 
\begin{equation}
\int{\rm d}\vec{v}G_{z}(\vec{v})=1
\end{equation}
is set for the eigenequation
\begin{equation}
{\mathcal K}G_{z}(\vec{v})=z G_{z}(\vec{v})
\end{equation} 
As the complex eigenvalues of ${\mathcal K_0}$ are 
determined by $\epsilon_0(z)=0$, the complex eigenvalues $z_j$ ($j=1,2,...,m$) 
of ${\mathcal K}$ are determined by $\epsilon(z)=0$ where 
\begin{equation}
\epsilon(z)=1+\pi\sum_n\int_0^{\infty}v_{\bot}{\rm d}v_{\bot}
\int_{-\infty}^{\infty}{\rm d}v_{\|}{J_n J_{n-1}\eta_{\bot}+J_n J_{n+1}
\eta_{\bot}+2J^2_n \eta_{\|}\over 2(z-m\omega_0-k_{\|}v_{\|})}
\end{equation}
Here and after, all summations are assumed to be over all integers. $J_n$'s are  
the Bessel functions $J_n (k_{\bot}v_{\bot}/\omega_0)$. Two functions, 
$\eta_{\bot}$ and $\eta_{\|}$, are defined as
\begin{equation}
\eta_{\bot}={\omega_p^2 \over k^2}k_{\bot}
{\partial\over\partial v_{\bot}}f_0(\vec{v})~~~,~~~~~~~~~~~
\eta_{\|}={\omega_p^2 \over k^2}k_{\|}
{\partial\over\partial v_{\|}}f_0(\vec{v})
\end{equation}
Note the equilibrium state function $f_0(\vec{v})$ has been assumed to 
be a function of only $v_{\bot}$ and $v_{\|}$, and $m$, the number of the 
discrete complex eigenvalues, is possible to be infinite.
The eigenfunction corresponding to $z_j$ is 
\begin{equation}
G_j(\vec{v})=\exp({\rm i}{k_{\bot}v_{\bot}\sin\theta\over\omega_0})\sum_n e^{-{\rm i}n\theta}
{(J_{n+1}+J_{n-1})\eta_{\bot}/2+J_n\eta_{\|}\over z_j-n\omega_0-k_{\|}v_{\|}}
\end{equation}
For a real eigenvalue $\mu$, we have
\begin{equation}
\begin{array}{rl}
G_{\mu}(\vec{v})=&\displaystyle 
\exp({\rm i}{k_{\bot}v_{\bot}\sin\theta\over\omega_0})\sum_n e^{-{\rm i}n\theta}{\mathcal P}
{(J_{n+1}+J_{n-1})\eta_{\bot}/2+J_n\eta_{\|}\over \mu-n\omega_0-k_{\|}v_{\|}}\\~\\
&\displaystyle +\sum_n a_n(v_{\|},v_{\bot})
\exp{-{\rm i}({k_{\bot}v_{\bot}\sin\theta\over\omega_0}-n\theta)}\delta(\mu-n\omega_0-k_{\|}v_{\|})
\end{array}
\end{equation}
where $a_n$'s are arbitrary functions as long as they satisfy the only constraint, the
normalization condition (30). This means that there are infinitely many eigenfunctions 
$G_{\mu}(\vec{v})$ corresponding to each real eigenvalue $\mu$. The infinite degeneracy 
arises again. It is easy to check that all these expressions reduce to the ones for 
$\vec{B}_0 =0$ . \\

We see that there are many similarities among ${\bar {\mathcal K}}_0$,  
${\mathcal K}_0$ and ${\mathcal K}$. But there is one important difference: 
the real eigenvalues of ${\bar {\mathcal K}}_0$
are not degenerate while the real eigenvalues of both ${\mathcal K}_0$ and 
${\mathcal K}$ are uncountably infinitely degenerate. As it has been shown by Case~\cite{Cas}, 
the solution of equation (13) can be expanded in terms of the van Kampen-Case modes.
In other words, we can use the eigenfunction expansion method to 
solve equation (13) just as we often do to solve 
the Schr{\" o}dinger equation. However, this method is not suitable for solving equations (10) 
and (11) even though we know the eigenfunctions of ${\mathcal K}_0$ and ${\mathcal K}$. 
Due to the infinite degeneracy, there is no obvious way that 
functions can be expanded in terms of these eigenfunctions. It is a common belief that 
for an initial value problem of a linear evolution system, to find the solution is
equivalent to find the eigenfunctions and spectrum of the linear operator
in this system. Seemingly, this is not the case for ${\mathcal K}_0$ and ${\mathcal K}$.

\section{Adjoint Equations and Operators}
Operator ${\mathcal K}$ is not self-adjoint, so it is interesting to 
know the functions orthogonal to
its eigenfunctions. For this purpose, we consider the adjoint equation of (10) 
\begin{equation}
{\rm i}{\partial\over \partial t}f_(\vec{v},t)={\tilde {\mathcal K}}f(\vec{v})
\end{equation}
where the linear operator ${\tilde {\mathcal K}}$ is defined as 
\begin{equation}
{\tilde {\mathcal K}}f(\vec{v})=\vec{k}\cdot\vec{v}f(\vec{v})+
{\rm i}{e\over m_e}(\vec{v}\times\vec{B}_{0})
\cdot{\partial\over \partial \vec{v}}f(\vec{v})-\int{\rm d}\vec{v}'\eta(\vec{v}')f(\vec{v}')
\end{equation}
The discrete complex eigenvalues of ${\tilde {\mathcal K}}$
are also determined by equation (32). Its eigenfunction corresponding to $z_j$ is 
\begin{equation}
{\tilde G}_j(\vec{v})=\exp({\rm i}{k_{\bot}v_{\bot}\sin\theta\over\omega_0})\sum_n e^{-{\rm i}n\theta}
{J_n\over z_j-n\omega_0-k_{\|}v_{\|}}
\end{equation}
while its eigenfunction of corresponding to a real eigenvalue $\mu$ is 
\begin{equation}
\begin{array}{rl}
{\tilde G}_{\mu}(\vec{v})=&\displaystyle 
\exp({\rm i}{k_{\bot}v_{\bot}\sin\theta\over\omega_0})\sum_n e^{-{\rm i}n\theta}
{\mathcal P}{J_n\over \mu-n\omega_0-k_{\|}v_{\|}}\\~\\
&\displaystyle +\sum_n {\tilde a}_n(v_{\|},v_{\bot})
\exp{-{\rm i}({k_{\bot}v_{\bot}\sin\theta\over\omega_0}-n\theta)}\delta(\mu-n\omega_0-k_{\|}v_{\|})
\end{array}
\end{equation}
where ${\tilde a}_n$'s are subject only to the normalization condition
\begin{equation}
\int{\rm d}\vec{v}\eta(\vec{v}){\tilde G}_{\mu}(\vec{v})=1
\end{equation}
It means that the real eigenvalues of ${\tilde {\mathcal K}}$ are also infinitely degenerate.
Therefore, we see that the adjoint operator ${\tilde {\mathcal K}}$ has the same 
spectrum structure and similar eigenfunctions as the original operator ${\mathcal K}$. 
By straightforward substitution, it can be proved
\begin{equation}
\int{\rm d}\vec{v}[{\tilde {\mathcal K}}f(\vec{v})]^* g(\vec{v})=
\int{\rm d}\vec{v}f^*(\vec{v})[{\mathcal K}g(\vec{v})]
\end{equation}
where ``$*$'' represents the complex conjugate. This relation leads easily to 
the orthogonal relation between the eigenfunctions of ${\tilde {\mathcal K}}$ and ${\mathcal K}$
\begin{equation}
(z^*-z')\int{\rm d}\vec{v}{\tilde G}^*_{z}(\vec{v})G_{z'}(\vec{v})=0 
\end{equation}\\
One interesting point to note is that for the complex eigenvalues we have
\begin{equation}
\int{\rm d}\vec{v}{\tilde G}^*_{z_j}(\vec{v})G_{z_j}(\vec{v})=0
\end{equation}
since $z_j^*\neq z_j$.
The adjoint equation of (11) can be considered as a special case of $\vec{B}_0=0$, no
additional treatment is necessary.\\
 
\section{Resolvent Method}
As it has been pointed out in the previous sections, the conventional eigenfunction 
expansion method is not suitable for solving equation (10). We have to resort to other methods. 
It turns out that the resolvent method is a successful choice as it is for the one dimensional 
equation (13)~\cite{Art}. The resolvent of ${\mathcal K}$ is defined as
\begin{equation}
{\mathcal R}(z)={1\over z-{\mathcal K}}
\end{equation}
where $z$ is a complex variable. 
If the initial function is $f(\vec{v},0)$ then the function at time $t$ is
\begin{equation}
f(\vec{v},t)={1\over 2\pi {\rm i}}\oint{\rm d}z e^{-{\rm i}zt}{\mathcal R}(z)f(\vec{v},0)
\end{equation}
where the integration contour surrounds all the singularities of ${\mathcal R}(z)$. 
For most operators, how their resolvents act on a function can be found only approximately.
In this case, the explicit form of resolvent is found exactly ( the detailed derivation is
given in the Appendix ), which is
\begin{equation}
{\mathcal R}(z)f(\vec{v},0)=F(z,\vec{v})-G(z,\vec{v})
{\int{\rm d}\vec{v}f(\vec{v},0){\tilde G}^*(z^*,\vec{v})\over \epsilon(z)}
\end{equation}
where
$$
\begin{array}{rl}
G(z,\vec{v})=&\displaystyle \exp({\rm i}{k_{\bot}v_{\bot}
\sin\theta\over\omega_0})\sum_n e^{-{\rm i}n\theta}
{(J_{n+1}+J_{n-1})\eta_{\bot}/2+J_n\eta_{\|}\over z-n\omega_0-k_{\|}v_{\|}}\\~\\

{\tilde G}(z,\vec{v})=&\displaystyle \exp({\rm i}{k_{\bot}v_{\bot}\sin\theta\over\omega_0})
\sum_n e^{-{\rm i}n\theta}{J_n\over z-n\omega_0-k_{\|}v_{\|}}\\~\\

F(z,\vec{v})=&\displaystyle \exp({\rm i}{k_{\bot}v_{\bot}\sin\theta\over\omega_0})
\sum_{m,n}f_m J_n e^{-{\rm i}(m+n)\theta}{1\over z-(m+n)\omega_0-k_{\|}v_{\|}}
\end{array}
$$
with
$$f(\vec{v},0)=\sum_m f_m(v_{\|},v_{\bot})e^{{\rm i}m\theta}$$
Similarly, for ${\tilde {\mathcal K}}$ we have
\begin{equation}
{1\over z-{\tilde {\mathcal K}}}f(\vec{v},0)=F(z,\vec{v})-{\tilde
G}(z,\vec{v})
{\int{\rm d}\vec{v}f(\vec{v},0) G^*(z^*,\vec{v})\over \epsilon(z)}
\end{equation}

As is well-known, the singularities of the resolvent ${\mathcal R}(z)$ give the 
spectrum of ${\mathcal K}$. It is not hard to see from (46) that the singularities of 
${\mathcal R}(z)$ include the zeroes of $\epsilon(z)$,
discrete real poles, $n\omega_0+k_{\|}v_{\|}$, and a branch cut along the real axis. 
Therefore the spectrum of ${\mathcal K}$ is just what we obtained earlier: all real values 
and a set of discrete complex values determined by $\epsilon(z)=0$. Let us write down 
the solution explicitly by choosing a specific contour
\begin{equation}
\begin{array}{rl}
f(\vec{v},t)=&\displaystyle -{1\over 2\pi {\rm i}}\int_{C_+}{\rm d}z 
e^{-{\rm i}zt}\{F(z,\vec{v})-G(z,\vec{v})
{\int{\rm d}\vec{v}f(\vec{v},0){\tilde G}^*(z^*,\vec{v})\over \epsilon(z)}\}\\~\\

&\displaystyle +{1\over 2\pi {\rm i}}\int_{C_-}{\rm d}z e^{-{\rm i}zt}\{F(z,\vec{v})-G(z,\vec{v})
{\int{\rm d}\vec{v}f(\vec{v},0){\tilde G}^*(z^*,\vec{v})\over \epsilon(z)}\}
\end{array}
\end{equation}
where $C_{\pm}$ run parallel to the real axis and are chosen such that all the 
singularities of ${\mathcal K}(z)$ are enclosed between them. 
We can use the residue theorem to calculate the above integration.
If we are only interested in the solution  for $t>0$, then 
two big semi-circles can be attached from below to $C_{\pm}$ to 
make two closed contours $O_{\pm}$. 
Since there are no singularities inside the contour $O_-$ the 
second part of the integration (48)
is identical to zero. In contrast,  $O_+$ encloses singularities: poles, such as $z_j$,
and a branch cut along the real axis. To get rid of
the troubling branch cut, we can replace the functions in the integrand with their plus 
analytic continuations \cite{Bal}. These plus continuations are identical to their 
original functions in the upper-half complex plane but are different in the lower-half
plane. They have no branch cuts but have new poles different from the original poles 
in the lower-half plane. These new poles are the 
mathematical origin of the Landau dampings~\cite{Bal, Ich} in this plasma model.\\
 
When $\vec{B}_0=0$, equation (48) becomes
\begin{equation}
{1\over z-{\mathcal K}_0}f(\vec{v},0)={f(\vec{v},0)\over z-\vec{k}\cdot\vec{v}}
-{\eta(\vec{v})\over z-\vec{k}\cdot\vec{v}}
{\int{\rm d}\vec{v}{f(\vec{v},0)\over z-\vec{k}\cdot\vec{v}}\over \epsilon_0(z)}
\end{equation} 
Therefore, the solution of equation (11) is 
\begin{equation}
\begin{array}{rl}
f(\vec{v},t)=&\displaystyle -{1\over 2\pi {\rm i}}\int_{C_+}{\rm d}z e^{-{\rm i}zt}
\{{f(\vec{v},0)\over z-\vec{k}\cdot\vec{v}}-{\eta(\vec{v})\over z-\vec{k}\cdot\vec{v}}
{\int{\rm d}\vec{v}{f(\vec{v},0)\over z-\vec{k}\cdot\vec{v}}\over \epsilon_0(z)}\}\\~\\

&\displaystyle +{1\over 2\pi {\rm i}}\int_{C_-}{\rm d}z e^{-{\rm i}zt}\{
{f(\vec{v},0)\over z-\vec{k}\cdot\vec{v}}-{\eta(\vec{v})\over z-\vec{k}\cdot\vec{v}}
{\int{\rm d}\vec{v}{f(\vec{v},0)\over z-\vec{k}\cdot\vec{v}}\over \epsilon_0(z)} \}
\end{array}
\end{equation}\\
After the integration over the two components of $\vec{v}$ perpendicular to $\vec{k}$, the
above equation can be reduce to one dimensional
\begin{equation}
\begin{array}{rl}
{\bar f}(v,t)=&\displaystyle -{1\over 2\pi {\rm i}}\int_{C_+}{\rm d}z e^{-{\rm i}zt}
\{{{\bar f}(v,0)\over z-kv}-{{\bar \eta}(v)\over z-kv}
{\int{\rm d}v{{\bar f}(v,0)\over z-kv}\over \epsilon_0(z)}\}\\~\\

&\displaystyle +{1\over 2\pi {\rm i}}\int_{C_-}{\rm d}z e^{-{\rm i}zt}\{
{{\bar f}(v,0)\over z-kv}-{{\bar \eta}(v)\over z-kv}
{\int{\rm d}v{{\bar f}(v,0)\over z-kv}\over \epsilon_0(z)} \}
\end{array}
\end{equation}\\
which is the solution of equation (13).\\
 
It is important to note here that the eigenfunctions ${\bar g}_z(v)$ 
of ${\bar {\mathcal K}}_0$ can be constructed through its resolvent 
solution (51). The way was demonstrated clearly in~\cite{Art}: Contour $C_{\pm}$
is divided into two parts, small circles surrounding poles and two
straight lines
approaching to the real axis from both below and above. The first part gives easily
the eigenfunctions ${\bar g}_i(v)$ corresponding to the complex eigenvalues.
The second part leads to the real eigenfunctions ${\bar g}_{\nu}(v)$.
For operators  ${\mathcal K}_0$ and ${\mathcal K}$, their complex 
eigenfunctions $g_j(\vec{v})$ and
$G_j(\vec{v})$ can be constructed easily similarly through their resolvent solutions 
(48) and (50), respectively. But straightforward calculation immediately
shows the 
technical difficulties for the real eigenfunctions 
$g_{\mu}(\vec{v})$ and $G_{\mu}(\vec{v})$ be constructed out of (48) and
(50)
in a similar way. It is not worthwhile to dwell on the technical difficulties which involve
very complicated calculations and formulas
since one may argue there may be another method to circumvent these difficulties. Let us focus
on the underlying reason, the uncountably infinite degeneracy.
Usually, the eigenfunctions of a linear operator can be labelled exactly either by a set of 
real numbers
or integers or both. For operators ${\mathcal K}_0$ and ${\mathcal K}$, I still labelled 
their eigenfunctions by real numbers and integers. But the labelling is not exact since there
are infinite number of eigenfunctions for each real number. It certainly requires the introduction
of non-trivial measure and other concepts in functional analysis to expand any function in the vector space 
expanded by the eigenfunctions $g_{\mu}(\vec{v})$ or $G_{\mu}(\vec{v})$. 
All of this is unlikely to be
achieved by some clever algebraic manipulations with (48) and (50).

\section{Conclusion}
In summary, I have studied the linearized Vlasov equation of plasma in an external constant
magnetic field $\vec{B}_0$ and found its exact solution by the resolvent method. Also the 
corresponding linear Vlasov operator ${\mathcal K}$ has been discussed and 
its eigenfunctions are found. The spectrum of this unbounded operator has two parts: 
continuous real eigenvalues and discrete complex eigenvalues. This is very 
similar to the spectrum of ${\bar {\mathcal K}}_0$, the one dimensional linear operator 
derived from the linearized Vlasov equation without external fields~\cite{van, Cas}. 
But there is a striking new feature for ${\mathcal K}$ that 
the real eigenvalues of ${\mathcal K}$ are uncountably infinitely degenerate. 
This new feature leads to the difficulty in 
expanding functions in terms of the eigenfunctions of ${\mathcal K}$.

\newpage
\begin{center}
{\bf \Large Acknowledgment}
\end{center}
I am very grateful to Mr. Suresh Subbiah for helpful discussions and Prof. Qian Niu for
his support. \\~\\~\\

\begin{center}
{\bf \Large Appendix}
\end{center}
In this appendix, the detailed derivation leading to (46) is given. The eigenfunctions  
 $G_j(\vec{v})$ and $G_{\mu}(\vec{v})$ can be found in the similar way. 
The method is called the integration along the characteristic curves ~\cite{Cou} or the integration
along the ``unperturbed orbits''~\cite{Ich}. In the following calculations, it is assumed that
Im$z>0$. To make formula compact, let us set two notations
$$
F = ( z - {\mathcal K} ) f(\vec{v})   \eqno(A.1)
$$
$$
E = \exp ({\rm i}\vec{k}\cdot\vec{x}-{\rm i}zt) \eqno(A.2)
$$
It is easy to verify with the aid of (9) that
$$
[{\rm i}{\partial\over \partial t} + {\rm i}\vec{v}\cdot{\partial\over \partial \vec{x}}
-{\rm i}{e \over m_e}(\vec{v}\times\vec{B_0})\cdot{\partial\over \partial \vec{v}}]
(fE) = FE - E\eta\int f \eqno(A.3)
$$
where $\eta = \eta(\vec{v})$, $f = f(\vec{v})$ and the non-specified integration is 
over velocity. Along the ``unperturbed orbit'', i.e., the orbit in the phase space
of one electron moving in magnetic field $\vec{B_0}$, we have
$$
{{\rm d}\over {\rm d}t} = {\partial\over \partial t} + \vec{v}\cdot
{\partial\over \partial\vec{x}} - {e \over m_e}(\vec{v}\times\vec{B_0})
\cdot{\partial\over \partial \vec{v}} \eqno(A.4)
$$
Then equation (A.3) becomes
$$
{\rm i}{{\rm d}\over {\rm d}t}(fE)=FE-\eta E\int f
\eqno(A.5)
$$
Integrating it, we get
$$
{\rm i}f=\int_{-\infty}^t{\rm d}t'F'E'E^{-1}-
(\int_{-\infty}^t{\rm d}t'\eta'E'E^{-1})\int f\eqno(A.6)
$$
where the primed functions mean the functions of primed variables. 
Note in the above integration over time $t$, the fact Im$z>0$ has been used to choose 
$-\infty$ as the lower limit.
Another integration with respect to velocity gives
$$
\int f = {\int\int_{-\infty}^t{\rm d}t'F'E'E^{-1}\over 
{\rm i} + \int\int_{-\infty}^t{\rm d}t'\eta'E'E^{-1}}
\eqno(A.7)
$$
Plugging this
back into (A.6), we have
$$
\begin{array}{ccl}
(z-{\mathcal K})^{-1}F&=&\displaystyle -{\rm i}
\int^t_{-\infty}{\rm d}t'F'E'E^{-1}\\~\\
&+&\displaystyle\int^t_{-\infty}{\rm d}t'\eta'E'E^{-1}
{\int\int_{-\infty}^t{\rm d}t'F'E'E^{-1}\over 
{\rm i} + \int\int_{-\infty}^t{\rm d}t'\eta'E'E^{-1}}
\end{array}
\eqno(A.8)
$$
Adopting the coordinate system and notations in (27) and (28), we write out
(A.8) explicitly
$$
\begin{array}{ccl}
{\mathcal R}(z)f(\vec{v})&=&\displaystyle -i\int_0^\infty{\rm d}\tau 
f(v_\|, v_\bot, \theta-\omega_0 t) e^{-{\rm i}\phi(\tau)}\\~\\
&+&\displaystyle \int_0^\infty{\rm d}\tau\eta(\vec{v}')e^{-{\rm i}\phi(\tau)}
{\int{\rm d}\vec{v}\int_0^\infty{\rm d}\tau
f(v_\|, v_\bot, \theta-\omega_0 t) e^{-{\rm i}\phi(\tau)}\over 
1 - {\rm i}\int{\rm d}\vec{v}\int_0^\infty{\rm d}\tau
\eta(\vec{v}') e^{-{\rm i}\phi(\tau)}}
\end{array}
\eqno(A.9)
$$
where
$$
\phi(\tau)={k_\bot v_\bot\over \omega_0}(\sin\theta-\sin(\theta-\omega_0\tau))
+(k_\|v_\|-z)\tau
\eqno(A.10)
$$
$$
\eta(\vec{v}')=\eta_\bot\cos(\theta-\omega_0\tau)+\eta_\|
~~~~~~~~~~~~~~~~~~~~~~~~~~~~\eqno(A.11)
$$
The case Im$z<0$ can be dealt with similarly, which also leads to (A.9). 
The step from (A.9) to (46) is direct integration with respect to $\tau$ using Bessel functions.


\newpage

\begin{thebibliography}{99}
\bibitem{Vla}
Vlasov, A., {\it J. Phys. USSR}, {\bf 9}(1945) 25

\bibitem{Lan}
Landau, L., {\it J. Phys. USSR}, {\bf 10}(1946) 26

\bibitem{Bal}
Balescu, R., {\it Statistical Mechanics of Charged Particles}, Interscience, New York (1963)

\bibitem{Ich}
Ichimaru, {\it Basic Principles of Plasma Physics}, Benjamin/Cummings, (1973)

\bibitem{Cas}
Case, K. M., {\it Ann. Phys.} {\bf 7} (1959) 349

\bibitem{Art}
Arthur, M.D. et al, {\it Phys. of Fluids}, {\bf 20} (1977) 1296

\bibitem{van}
Van Kampen, N.G., {\it Physica}, {\bf 21}(1955) 949; {\bf 23} (1957) 641

\bibitem{Bac}
Backus, G., {\it J. of Math. Phys.}, {\bf 1} (1960) 178

\bibitem{Cou}
Courant \& Hilbert, {\it Methods of Mathematical Physics}(II), Interscience, New York, (1962) 

\end{thebibliography}
\end{document}